\newcommand{\half}{\frac{1}{2}}
\newcommand{\adag}{a^{\dag}}

\newcommand{\ds}{\displaystyle}
 \documentstyle[preprint,aps,tighten]{revtex}
\begin{document}
\draft
\title{Generalized Heisenberg relation and Quantum Harmonic
Oscillators}
\author{ P. Narayana Swamy }
\address{Department of Physics, Southern Illinois University,
Edwardsville IL 62026}

\maketitle
\begin{abstract}
\end{abstract}

We study  the consequences of the generalized Heisenberg
uncertainty relation which admits a minimal uncertainty in length
such as the case in a theory of quantum gravity. In particular,
 the theory of quantum harmonic oscillators  arising
from such a generalized uncertainty relation is examined. We
demonstrate that all the standard properties of the quantum
harmonic oscillators prevail when we employ a generalized
momentum. We also show that quantum electrodynamics and coherent
photon states can be described in the familiar standard manner
despite the generalized uncertainty principle.

 \vspace{2.8in}
Electronic address: pswamy@siue.edu \vspace{.2in}

  \pacs{PACS $02.20. Sv,\; 02.40. Gh,\;  03.65. Ca,\; 03.65.-w,\; 03.70+ k$}

\section{Introduction}
We begin with the generalized commutation relation,
\begin{equation}\label{1a}
    [q_{{\bf k},i}, p_{{\bf k}',j}]= i \hbar
    \delta_{{\bf k},{\bf k}'}\,
    \delta_{i,j}\left (1 + \beta p_{{\bf k},i}\, p_{{\bf k}',j}
     \right )\, ,
\end{equation}
which reduces to the fundamentally simple form if we consider a
single degree of freedom,
\begin{equation}\label{1}
    [q,p]= i \hbar (1 + \beta p^2)\, .
\end{equation}
This  leads to the generalized Heisenberg uncertainty relation
\begin{equation}\label{2}
    \Delta q \Delta p \geq \half \hbar \left( 1  + \beta (\Delta
    p)^2 \right )\, ,
\end{equation}
where $\beta$ is positive and independent of $\Delta q, \Delta p$.
The above relation implies \cite{Kempf} a minimum length described
 by the
parameter $\beta$. This type of generalized uncertainty relation
has appeared in the context of quantum gravity and string theory
and  is based on the premise that a minimal length should quantum
theoretically be described as a minimal uncertainty in position
measurements. Several consequences of the above relations have
been of much interest \cite{Kempf,Chang,Camacho,Xiang,Goldin}
recently. This has lead to a great deal of interest in several
aspects such as the Hilbert space structure of the theory in the
presence of quantum gravity, the formulation of Liouville theorem
in statistical physics, the cosmological constant problem, the
deformation of the black body radiation,  consequences in quantum
electrodynamics in the presence of quantum gravity and deformed
local current algebra compatible with a fundamental length scale.
We shall, for the sake of expediency, regard the generalized
commutation relation as due to a deformation, as a result of
quantum gravity and accordingly refer to the system obeying
Eq.(\ref{2}) as deformed; we shall refer to the limit $\beta
\rightarrow 0$ simply as the undeformed case. Our goal here is a
modest investigation of the question: what are the consequences
when the physical system is viewed as a set of quantum harmonic
oscillators?

 We shall consider single level
oscillators, for convenience. The paper is organized as follows.
Section I begins with a summary of known results  and formulates
the algebra and the Fock space of the quantum oscillators in the
presence of deformation. The ground state is determined in Section
II. We study the Heisenberg equations of motion in Section III and
demonstrate in particular that Liouville theorem prevails in the
standard form despite the presence of deformation when the system
is described by generalized momenta. Section IV is devoted to
summarizing the results in quantum electrodynamics and quantum
optics to show that consequences of deformation leading to a
generalized Heisenberg uncertainty principle are not explicitly
present. Section V concludes with a summary.

Let us proceed by introducing the harmonic oscillators by means of
the operators,
\begin{equation}\label{3}
    a= \sqrt{\frac{m \omega}{2 \hbar}}\left ( q +
    \frac{i P}{m \omega}\right )\,
    \quad
\adag= \sqrt{\frac{m \omega}{2 \hbar}}\left (q - \frac{i P}{m
\omega}\right )\,,
\end{equation}
where $q$ is the coordinate and  $P$ is the generalized momentum
whose properties will be described below. We can choose
\begin{equation}\label{4}
    P = p + f(p)\, ,
\end{equation}
where $f(p)$ is a general function which must be so chosen in
order that it will lead to simple and desirable properties for the
system of harmonic oscillators. We may in fact, choose,
\begin{equation}\label{5}
    f(p)= \sum_{r=1}^{\infty}\; \frac{(-\beta)^r}{2 r +1}\; p^{2r
    +1}\,
\end{equation}
for this reason. We shall adhere to the suggestion of  Camacho
\cite{Camacho} who has shown that this choice leads to the
uncertainty relation in Eq.(\ref{2}), as well as to the further
useful result
\begin{equation}\label{6}
    [q,f(p)]= - i \hbar \beta p^2\, ,
\end{equation}
and consequently we obtain
\begin{equation}\label{7}
    [q,P]= i \hbar\, .
\end{equation}
In other words, the choice
\begin{equation}\label{8}
    P= p + f(p) = \frac{1}{\sqrt{\beta}} \arctan (\sqrt{\beta} p)\, ,
\end{equation}
determines  $P $ so that it plays  the role of a generalized
momentum in the sense of the canonical commutation relation,
Eq.(\ref{7}). Consequently, from the representation in
Eq.(\ref{3}), we obtain
\begin{equation}\label{9}
    a \adag = \frac{1}{\hbar \omega}\, H + \half, \quad
    \adag a = \frac{1}{\hbar \omega}\, H - \half\, ,
\end{equation}
where
\begin{equation}\label{10}
    H= \half m \omega^2 q^2 + \frac{P^2}{2m}
\end{equation}
is the Hamiltonian in terms of the generalized momentum $P$. We
immediately obtain the result
\begin{equation}\label{11}
    a \adag - \adag a = 1\, .
\end{equation}
This is the standard algebra of oscillators, a remarkable result
\cite{Camacho} despite the deformation introduced as in
Eq.(\ref{2}). In other words, the standard algebra of harmonic
oscillators prevails despite the deformation due to quantum
gravity. To develop this idea further, we proceed as follows. We
build the Fock states from the ground state by the construction
\begin{equation}\label{14}
    |n\rangle = \; \frac{(\adag)^n}{\sqrt{n !}}\; |0\rangle\, .
\end{equation}
Let us define the operator relation $\adag a = N$, the number
operator which in turn leads to $a \adag = N+1$ as a consequence
of the algebra above. The number operator acts on the states so
that $N |n \rangle = n |n \rangle$, defining the Fock states
$|n\rangle$, where $n=0,1,2, \cdots\; $. The operators $a, \adag$
are seen to satisfy the relations
\begin{equation}\label{12}
    [N,a]=-a, \quad [N,\adag]= \adag\,.
\end{equation}
By examining $N\adag |n\rangle$, we conclude in the standard
manner that $\adag$ is the creation operator and $a$ is the
annihilation operator.  Indeed we find,
\begin{equation}\label{13}
    a|n\rangle = \sqrt{n}\, |n-1\rangle\, , \quad
\adag|n\rangle = \sqrt{n+1}\, |n+1\rangle\, ,
\end{equation}
which is no different from the standard results of the undeformed
oscillator. All this agrees with the well-known properties of the
operators for the standard system of quantum harmonic oscillators.
What is significant now is that these results are valid for the
case when the Heisenberg commutation relation is generalized to
identify a fundamental length in the uncertainty principle arising
from a deformation such as in quantum gravity. From Eq.(\ref{10}),
we further obtain the important relation
\begin{equation}\label{15}
    H=\half \hbar \omega (\adag a + a \adag)= \hbar \omega
    (N+\half)\, ,
\end{equation}
and hence the energy spectrum is given by
\begin{equation}\label{16}
    E_n=\hbar \omega
    (n+\half)\, ,
\end{equation}
when $n=0,1, \, \cdots \,$, for the quantum system with the
generalized Heisenberg relation when the Hamiltonian is given by
Eq.(\ref{11}). Despite the introduction of a fundamental length,
the standard properties of the harmonic oscillators thus prevail,
and so does the energy spectrum. Here the generalized momentum may
be expressed in the form of a series,
\begin{equation}\label{17}
    P= \frac{1}{\sqrt{\beta}} \arctan (\sqrt{\beta} p) =
    \frac{1}{\sqrt{\beta}}\left \{\sqrt{\beta} p- \frac{(\sqrt{\beta}
    p)^3}{3}+\frac{(\sqrt{\beta}
    p)^5}{5} + \cdots
     \right \}
\end{equation}
when $\beta \neq 0$, and we recover the ordinary momentum in the
limiting case:
\begin{equation}\label{18}
    \lim_{\beta \longrightarrow 0}\, P=p\, .
\end{equation}
 The square of the momentum occurring in the
Hamiltonian may be identified by the series
\begin{equation}\label{19}
    P^2= p^2- \frac{2 \beta p^4}{3}+ \frac{23\beta^2 p^6}{45}-
    \frac{44 \beta^3 p^8}{105}+\frac{563 \beta^4 p^{10}}{1575}
    + \cdots\, .
\end{equation}
We expect that this generalized momentum is related to the
coordinate in the form  $P=-i \hbar (\partial/ \partial_q)$.
Indeed this relation is true up to a constant phase as pointed out
in Dirac's textbook \cite{Dirac,referee}. This introduces a
tremendous simplification in the theory. We may consider the
ground state, defined by
\begin{equation}\label{20}
    a \psi_0 = (q + \frac{i P}{m \omega})\; \psi_0 =0\, ,
\end{equation}
in terms of the coordinate $x$ and the generalized momentum $P$.
This leads to the standard harmonic oscillator with the solution
\begin{equation}\label{21}
    \psi_0= (\frac{m \omega}{\pi \hbar})^{1/4}\; e^{-\ds m \omega ^2/2
    \hbar}\, .
\end{equation}
The other states can similarly be shown to correspond to the case
of the standard harmonic oscillators. It must be pointed out that
the Hamiltonian in Eq.(\ref{10}) involves the momentum $P$ which
has lead to the great simplification. Thus although the form of
the Hamiltonian is the same, the harmonic oscillator considered
here differs from that of references \cite{Kempf} and
\cite{Chang}.

\section{Equations of motion}

We shall now investigate the Heisenberg equations of motion for
operators in the form,
\begin{equation}\label{27a}
i \hbar \dot{F}=[F,H]\, ,
\end{equation}
 where $\dot{F}$ is the full time derivative of the operator $F$\,
 and we shall assume no explicit time dependence.
 In particular , let us consider the relation
\begin{equation}\label{28}
    i \hbar \dot{a}= [a,H]= \sqrt{\frac{m \omega}{2 \hbar}}
    \; \left [ \; q + \frac{i P}{m \omega}\, ,\,  H \right ]\, .
\end{equation}
We can evaluate the commutator $[a,H]$ either directly or by
making use of the form in Eq.(\ref{15}) and the result is
\begin{equation}\label{29}
\dot{a} = - i \omega a\, .
\end{equation}
Hence we recover the familiar form of time dependence of the
annihilation operator in a self-consistent manner:
\begin{equation}\label{30}
    a(t)=a_0 e^{- i \omega t}\, .
\end{equation}
Let us next consider
\begin{equation}\label{31}
 i \hbar \dot{p}= [p,H]\, .
\end{equation}
Evaluating in a straight forward manner, we obtain
\begin{equation}\label{32}
\dot{p}= - m \omega^2 (1 + \beta p^2) q \, ,
\end{equation}
which reduces to the standard result in the limit when $\beta$
vanishes. Furthermore,  we obtain
\begin{equation}\label{33}
    \dot{P}= - m\omega^2 q\, ,
\end{equation}
a more interesting result for the generalized momentum. Finally,
from
\begin{equation}\label{34}
    i \hbar \dot{q}= [q, H]\, ,
\end{equation}
we obtain the result
\begin{equation}\label{35}
    \dot{q}= \frac{P}{m}= \frac{1}{m \sqrt{\beta}}\, \arctan (\sqrt{\beta}
    p)\, ,
\end{equation}
which possesses the property $\lim_{\beta \longrightarrow 0}\;
\dot{q}= p/m$.  This can be expressed as a series in powers of
$\beta$. However, we may prefer the direct form in terms of the
 the generalized momentum $P$: this
 formulation seems  more direct and simple when expressed in terms of
 the generalized momentum.

It can be further verified that the Hamilton's equations are valid
in their canonical form:
\begin{equation}\label{35a}
    \dot{q}_i= -\frac{i}{\hbar}\; [q_i, H]=\partial_{P_i}\, H\, \quad
\dot{P}_i= - \partial_{q_i}\, H\,.
\end{equation}
Accordingly, the momentum $P$ is related to $p$ by a canonical
transformation.  In this form, it is interesting to observe the
quantum analog of the classical equations of motion, in particular
\cite{Schiff} the correspondence:
\begin{equation}\label{35b}
    \{A,\, B \}\;\Longrightarrow \; - \frac{i}{\hbar}[ A,\, B ]
    \, ,
\end{equation}
where the braces represent the Poisson bracket
\begin{equation}\label{35bb}
    \{A,B\}= \sum_{i=1}^{3N}\; \left (\frac{\partial A}{\partial
    q_i}\, \frac{\partial B}{\partial P_i}-
    \frac{\partial B}{\partial
    q_i}\, \frac{\partial A}{\partial P_i}\, ,
    \right )
\end{equation}
in terms of  $q_i$ the coordinate  and  $P_i$ the generalized
momentum.

 We shall now derive
the Liouville theorem of statistical mechanics and convince
ourselves that it is valid in the standard form even in the
deformed case when expressed in terms of the momentum $P$. Thus,
beginning from the continuity relation valid in fluid mechanics,
\begin{equation}\label{35c}
\int_{\omega} \;  d \omega\; \left ( \frac{\partial}{\partial
t}\rho (q_i, P_i) + \nabla \cdot \rho \, {\bf v}\right ) = 0\, ,
\end{equation}
where $\rho(q_i,P_i)$ is the phase space density in the space of
$3N$ dimensions, $i=1, \; \cdots \; 3N$ where the phase space is
made of the coordinates and the generalized momenta. We then
express the flux rate in the form
\begin{equation}\label{35d}
    \nabla \cdot \rho \, {\bf v}=
    \sum_{i=1}^{3N}\; \left \{\frac{\partial \rho}{\partial
    q_i}\,\dot{q}_i + \, \frac{\partial \rho}{\partial
    P_i}\,\dot{P}_i\,  + \rho \, \frac{\partial
    \dot{q}_i}{\partial q_i}\,  +
    \rho \, \frac{\partial
    \dot{P}_i}{\partial P_i}
    \right \}\, ,
\end{equation}
and accordingly,
\begin{equation}\label{35s}
\nabla \cdot \rho \, {\bf v}=\{ \rho, \, H \}
\end{equation}
 We may now employ the Hamilton equations, and
accordingly arrive at the result,
\begin{equation}\label{35e}
    \frac{d \rho(q_i, P_i)}{d t}= \frac{\partial \rho}{\partial
    t}+ \{\rho, \, H \}=
    0\, ,
\end{equation}
i.e.,  the total time derivative of the phase space density
vanishes as the system evolves along a phase space trajectory,
which is Liouville theorem in phase space. Consequently, the
theorem prevails in its standard form even in the deformed case,
such as when the deformation arises from quantum gravity.

\section{Photons}

In the case of quantum electrodynamics in the presence of
deformation, the Hamiltonian would be given by
\begin{equation}\label{36}
    H= \int\; d^3r \left (2 \pi c^2 |P|^2 + \frac{1}{8 \pi}
    (\nabla \times {\bf A})^2  \right )\, ,
\end{equation}
where the momentum conjugate to the vector potential is the
generalized momentum. The standard analysis then leads to the
Hamiltonian of the form
\begin{equation}\label{37}
    H= \sum^{'}_{{\bf k},\lambda}\left
    (4 \pi c^2 P^{\dag}_{{\bf k}, \lambda}
    P_{{\bf k}, \lambda} +\frac{k^2}{4 \pi}q^{\dag}_{{\bf k},\lambda}
    q_{{\bf k},\lambda}\right )\, ,
\end{equation}
where the prime on the sum indicates that the sum is over only
half the ${\bf k}$-space \cite{Schiff}. We may then introduce the
operators in the familiar manner\cite{Schiff},
\begin{equation}\label{38}
    a_{{\bf k},\lambda}=\half \{q_{{\bf k}, \lambda}
    + \frac{4 \pi i c}{k}P_{{\bf k},\lambda}\}\, e^{ikct}\, \quad
a^{\dag}_{{\bf k},\lambda}=\half \{q_{{\bf k}, \lambda}
    - \frac{4 \pi i c}{k}P_{{\bf k},\lambda}\}\, e^{-ikct}\,
\end{equation}
and the corresponding relations for the hermitian conjugate of the
above. We identify, in the standard manner, the number operator
\begin{equation}\label{39}
N_{{\bf k},\lambda}= \frac{k}{2 \pi \hbar c}\adag_{{\bf
k},\lambda}\,a_{{\bf k},\lambda}\,.
\end{equation}
In this manner, we may proceed to describe quantum electrodynamics
in terms of the harmonic oscillators, using the variables $q_{{\bf
k}, \lambda}$ and the generalized momenta $P_{{\bf k}, \lambda}$.
We proceed through the standard analysis and arrive at the final
result for the Hamiltonian,
\begin{equation}\label{40}
    H= \sum_{{\bf k},\lambda} \hbar c k (N_{{\bf k},\lambda}+
    \half)\, ,
\end{equation}
describing the quanta of electrodynamics.

Proceeding further, we observe that the theory of coherent photon
states \cite{Barnett} is of great interest and it is worthwhile
investigating the consequences of quantum gravity for coherent
states, in terms of the generalized uncertainty principle. For
this purpose, we shall consider single level modes for the sake of
simplicity in notation. The coherent states are defined by
\begin{equation}\label{41}
    |\alpha \rangle = e^{-\half |\alpha|^2} \sum_{n=0}^{\infty}\;
    \frac{\alpha^n}{\sqrt{n!}}|n\rangle\, ,
\end{equation}
which satisfy the property, $a|\alpha \rangle = \alpha |\alpha
\rangle$, where $a$ is the annihilation operator defined in
Eq.(\ref{38}). The probability that a coherent state contains the
$n$-quantum state is given by
\begin{equation}\label{41a}
    P(n)=\left | \langle n | \alpha \rangle \right |^2
    =e^{-\ds |\alpha|^2} \frac{|\alpha|^{2n}}{n!}\, .
\end{equation}
Instead of the usual orthogonality relation, the coherent states
obey the relation
\begin{equation}\label{41b}
    \langle \alpha |\beta \rangle =
    e^{ - \half \ds ( |\alpha|^2 + |\beta|^2 -
    2 \alpha^{\*}\beta)}
\end{equation}
 Employing the quadrature operator,
\begin{equation}\label{42}
    x_{\lambda}= \frac{1}{\sqrt{2}}\, \left (
    a\, e^{-i \lambda} + \adag e^{i \lambda}\right )\, ,
\end{equation}
we obtain  the result
\begin{equation}\label{43}
    \langle \alpha | x^2_{\lambda}|\alpha \rangle =
    \half \left ( \alpha^2 e^{-2i \lambda}+
    {\alpha^{\*}}^2e^{2i\lambda}
    + 2 |\alpha|^2 + 1\, ,
      \right )
\end{equation}
for the deformed case, just as in the standard undeformed case
\cite{Barnett}. In this manner, we obtain the familiar result for
the variance,
\begin{equation}\label{44}
    \Delta q_{\lambda} \; \Delta q_{\lambda + \pi/2} = \half.
\end{equation}
We accordingly find that  the theory of coherent states when the
generalized Heisenberg uncertainty principle prevails, is indeed
described by the properties of the standard theory with no
explicit consequences of the deformation.

\section{Summary}

We have studied the consequences of the generalized commutation
relation leading to the generalized Heisenberg uncertainty
relation, for a system described by quantum harmonic oscillators,
referring to this case as a deformation described by the parameter
$\beta$. Based on the fact that despite the deformation,  the
algebra of the creation and annihilation operators is no different
from the standard undeformed algebra requires that the theory be
described by a generalized momentum. We are thus able to describe
the oscillators in a self-consistent formulation. We are able to
show that the energy spectrum of the oscillators of a deformed
system is no different from that of the standard quantum
oscillators if the Hamiltonian is a quadratic sum of $q$ and $P$.
We have also studied the equations of motion in the deformed case,
in order to point out how they differ from the standard equations
of motion, and our results differ from that of ref.(\cite{Kempf}).
We employed the canonical Hamilton equations in order to derive
the Liouville theorem, namely that the phase space density is a
constant in time as the statistical system evolves. We have
demonstrated that in contrast to the discussion in
ref.(\cite{Chang}), the standard form of the Liouville theorem
prevails in the deformed case when the canonical variables are
described in terms of the generalized momentum. Finally, we have
been able to study quantum electrodynamics and the theory of
coherent state of photons to again show that there is no explicit
consequence of quantum gravity related fundamental length in these
cases.

We must point out the intimate connection between the undeformed
harmonic oscillator and the generalized uncertainty relation. Not
only is the oscillator algebra preserved by the generalization or
deformation of the uncertainty relation, the energy spectrum is
the same as of the standard form, and the ground state and excited
state wave functions of the harmonic oscillator can be constructed
in a straightforward manner. The generalized momentum $P$ defined
in Eq.(\ref{8}) is related to the momentum $p$ by a canonical
transformation as revealed by Eqs.(\ref{33}-\ref{36}). We might
thus regard the parameter $\beta$ as transformed away by a
canonical transformation.

 The main purpose of this work is to
address the question: can we reconcile the generalized uncertainty
relation with the requirement that the oscillator algebra be
preserved in its standard form? We are able to show that this can
be accomplished if we introduce a generalized momentum. By
extending the work of Camacho \cite{Camacho} in this manner, we
have been able to demonstrate that the form of the generalized
momentum is consistent with the generalized uncertainty relation
incorporating a minimal length as well as with the standard form
of the harmonic oscillator algebra.

\end{document}